\documentclass[aps,prb,reprint,groupedaddress]{revtex4-1}
\usepackage{graphicx}

\begin{document}


\title{Retarded interactions in Graphene systems}


\author{Bo E. Sernelius}
\email[]{bos@ifm.liu.se}
\affiliation{Division of Theory and Modeling, Department of Physics, Chemistry and Biology, Link\"{o}ping University, SE-581 83 Link{\"o}ping, Sweden}


\date{\today}

\begin{abstract}
We first demonstrate how two-dimensional sheets are incorporated in the formalism for planar structures. Then we derive the interaction in the geometry of two free-standing graphene sheets and of one graphene sheet above a substrate. Numerical results are produced for the fully retarded interaction at 0 K and at room temperature for undoped and doped graphene. Additional results are given both for a gold substrate and for an ideal metal substrate.

\end{abstract}


\maketitle
\section{Introduction\label{intro}}
The van der Waals (vdW) and Casimir interactions can be derived in many different ways. 
One way is to derive the interaction in terms of the electromagnetic normal modes\cite{Ser0} of the system. For planar structures the interaction energy per unit area can be written as
\begin{equation}
E = \hbar \int {\frac{{{d^2}k}}{{{{\left( {2\pi } \right)}^2}}}} \int\limits_0^\infty  {\frac{{d\omega }}{{2\pi }}} \ln \left[ {{f_k}\left( {i\omega } \right)} \right],
\label{equ1}
\end{equation}
where ${f_k}\left( {{\omega _k}} \right) = 0$ is the condition for electromagnetic normal modes. Eq.\,(\ref{equ1}) is valid for zero temperature and the interaction energy is the internal energy. At finite temperature the interaction energy is Helmholtz' free energy and can be written as
\begin{equation}
\begin{array}{l}
E = \frac{1}{\beta }\int {\frac{{{d^2}k}}{{{{\left( {2\pi } \right)}^2}}}} \sum\limits_{n = 0}^\infty {'} {\ln \left[ {{f_k}\left( {i{\omega _n}} \right)} \right];} \\
{\omega _n} = \frac{{2\pi n}}{{\hbar \beta }};\;n = 0,\,1,\,2,\, \ldots ,
\end{array}
\label{equ2}
\end{equation}
where $\beta  = 1/{k_B}T$. The integral over frequency has been replaced by a summation over discrete frequencies, so called Matsubara frequencies.
The prime on the summation sign indicates that the $n = 0$ term should be divided by two. 
 For planar structures the quantum number that characterizes the normal modes is {\bf k}, the two-dimensional (2D) wave vector in the plane of the interfaces.
The normal modes can be found from self-sustained charge and current densities, from self-sustained potentials or from self-sustained fields. Here, we will find self-sustained fields. To do that one solves the Maxwell equations (ME) in all regions of the geometry and makes use of the standard boundary conditions at all  interfaces between the regions. For the present task we need a geometry consisting of three regions and two interfaces, $1|2|3$. This geometry gives the mode condition function
\begin{equation}
{f_k} = 1 - {e^{ - 2{\gamma _2}kd}}{r_{21}}{r_{23}},
\label{equ3}
\end{equation}
where ${r_{ij}}$ is the amplitude reflection coefficient for a wave impinging on the interface between medium $i$ and $j$ from the $i$ side, $d$ is the thickness of region $2$ and ${\gamma _i} = \sqrt {1 - {{\tilde \varepsilon }_i}\left( \omega  \right){{\left( {\omega /ck} \right)}^2}} $. The function ${{{\tilde \varepsilon }_i}\left( \omega  \right)}$ is the dielectric function of medium $i$ and $c$ is the speed of light in vacuum.
For planar structures there are two types of mode, transverse magnetic (TM) or p-polarized and transverse electric (TE) or s-polarized. They have different amplitude reflection coefficients. At an interface where there are no 2D sheet the TM and TE amplitude reflection coefficients are
\begin{equation}
r_{ij}^{TM} = \frac{{{{\tilde \varepsilon }_j}{\gamma _i} - {{\tilde \varepsilon }_i}{\gamma _j}}}{{{{\tilde \varepsilon }_j}{\gamma _i} + {{\tilde \varepsilon }_i}{\gamma _j}}},
\label{equ4}
\end{equation}
and
\begin{equation}
r_{ij}^{TE} = \frac{{\left( {{\gamma _i} - {\gamma _j}} \right)}}{{\left( {{\gamma _i} + {\gamma _j}} \right)}},
\label{equ5}
\end{equation}
respectively.  Note that ${r_{ji}} =  - {r_{ij}}$ holds for both mode types. 

In next section we show how these amplitude reflection coefficients are modified at an interface containing a 2D sheet. In Sec.\,\ref{dielfunc} we discuss the dielectric function of undoped and doped graphene. Then follows the analytical expressions and numerical results for free-standing graphene sheets, Sec. Sec.\,\ref{2D sheets}, a graphene sheet above a gold substrate, Sec.\,\ref{gold}, and a graphene sheet above an ideal metal substrate,  Sec.\,\ref{ideal}. We end with summary and conclusions in Sec.\,\ref{summary}.

\section{How to incorporate 2D sheets in the formalism\label{2D sheets}}
There are different formulations of electromagnetism in the literature. The difference lies in how the conduction carriers are treated. In one formulation these carriers are lumped together with the external charges to form the group of free charges. Then only the bound charges contribute to the screening. We want to be able to treat geometries with metallic regions. Then this formulation is not suitable. In the formulation we use the conduction carriers are treated on the same footing as the bound charges. Thus, both bound and conduction charges contribute to the dielectric function. In the two formalisms the {\bf E} and {\bf B} fields, the true fields, of course are the same. However, the auxiliary fields  the {\bf D} and {\bf H} fields are different. To indicate that we use this alternative formulation we put a tilde above the {\bf D} and {\bf H} fields and also above the dielectric functions.

With this formulation MEs have the form
\begin{equation}
\begin{array}{l}
\nabla  \cdot {\bf{\tilde D}} = 4\pi {\rho _{ext}}\\
\nabla  \cdot {\bf{B}} = 0\\
\nabla  \times {\bf{E}} + \frac{1}{c}\frac{{\partial {\bf{B}}}}{{\partial t}} = 0\\
\nabla  \times {\bf{\tilde H}} - \frac{1}{c}\frac{{\partial {\bf{\tilde D}}}}{{\partial t}} = \frac{{4\pi }}{c}{{\bf{J}}_{ext}},
\end{array}
\label{equ6}
\end{equation}
and the boundary conditions at an interface between two media $1$ and $2$ are
\begin{equation}
\begin{array}{l}
\left( {{{\bf{E}}_2} - {{\bf{E}}_1}} \right) \times {\bf{n}} = 0\\
\left( {{{{\bf{\tilde D}}}_2} - {{{\bf{\tilde D}}}_1}} \right) \cdot {\bf{n}} = 4\pi {\left( {{\rho _s}} \right)_{ext}}\\
\left( {{{\bf{B}}_2} - {{\bf{B}}_1}} \right) \cdot {\bf{n}} = 0\\
\left( {{{{\bf{\tilde H}}}_2} - {{{\bf{\tilde H}}}_1}} \right) \times {\bf{n}} =  - \frac{{4\pi }}{c}{{\bf{K}}_{ext}}.
\end{array}
\label{equ7}
\end{equation}
We note that the sources to the fields in MEs are the external charge and current densities and also in the boundary conditions discontinuities in the normal component of the {\bf D} fields and tangential component of the {\bf H} fields are caused by external surface charge densities and external surface current densities, respectively. The unit vector {\bf n} is the surface normal pointing into region $2$.

The amplitude reflection coefficient gets modified if there is a 2D layer at the interface. We treat the 2D layer at the interface as external to our system. The fields will then induce external surface charge and current densities at the interface. Two of the boundary conditions are enough to get the modified Fresnel coefficients. The other two are redundant. We choose the following two
\begin{equation}
\begin{array}{l}
\left( {{{\bf{E}}_2} - {{\bf{E}}_1}} \right) \times {\bf{n}} = 0\\
\left( {{{{\bf{\tilde H}}}_2} - {{{\bf{\tilde H}}}_1}} \right) \times {\bf{n}} =  - \frac{{4\pi }}{c}{{\bf{K}}_{ext}} =  - \frac{{4\pi }}{c}\sigma {\bf{n}} \times \left( {{\bf{E}} \times {\bf{n}}} \right),
\end{array}
\label{equ8}
\end{equation}
where $\sigma$ is the conductivity of the 2D sheet. The modified amplitude reflection coefficient for a TM mode is
\begin{equation}
r_{ij}^{TM} = \frac{{{{\tilde \varepsilon }_j}{\gamma _i} - {{\tilde \varepsilon }_i}{\gamma _j} + 2{\gamma _i}{\gamma _j}{\alpha ^\parallel }}}{{{{\tilde \varepsilon }_j}{\gamma _i} + {{\tilde \varepsilon }_i}{\gamma _j} + 2{\gamma _i}{\gamma _j}{\alpha ^\parallel }}},
\label{equ9}
\end{equation}
where the polarizability of the 2D sheet is obtained from the dynamical conductivity
\begin{equation}
{\alpha ^\parallel }\left( {k,\omega } \right) = \frac{{2\pi i{\sigma ^\parallel }\left( {k,\omega } \right)k}}{\omega },
\label{equ10}
\end{equation}
and the dielectric function is
\begin{equation}
{\varepsilon ^\parallel }\left( {k,\omega } \right) = 1 + {\alpha ^\parallel }\left( {k,\omega } \right).
\label{equ11}
\end{equation}
For TM modes the tangential component of the electric field, which will induce the external current, is parallel to {\bf k}, so the longitudinal 2D dielectric function of the layer enters. The bound charges in the 2D sheet also contribute to the dynamical conductivity and the polarizability.

The modified amplitude reflection coefficient for a TE mode is
\begin{equation}
r_{ij}^{TE} = \frac{{{\gamma _i} - {\gamma _j} + 2{{\left( {\omega /ck} \right)}^2}{\alpha ^ \bot }}}{{{\gamma _i} + {\gamma _j} - 2{{\left( {\omega /ck} \right)}^2}{\alpha ^ \bot }}},
\label{equ12}
\end{equation}
where the polarizability of the 2D sheet is obtained from the dynamical conductivity
\begin{equation}
{\alpha ^ \bot }\left( {k,\omega } \right) = \frac{{2\pi i{\sigma ^ \bot }\left( {k,\omega } \right)k}}{\omega },
\label{equ13}
\end{equation}
and the dielectric function is
\begin{equation}
{\varepsilon ^ \bot }\left( {k,\omega } \right) = 1 + {\alpha ^ \bot }\left( {k,\omega } \right).
\label{equ14}
\end{equation}
For a TE wave the electric field is perpendicular to {\bf k}, so the transverse 2D dielectric function of the layer enters. 
The bound charges in the 2D sheet also contribute to the dynamical conductivity and the polarizability.

Now, when we have geometrical structures with 2D layers at some of the interfaces we just substitute the new reflection coefficients at the proper interfaces. Note that ${r_{21}} =  - {r_{12}}$ no longer applies so one has to be careful when making the substitutions. To arrive at the starting expression one might have used this relation. For the three layer system $1|2|3$ we find the most general mode condition functions as
\begin{equation}
\begin{array}{l}
f = 1 - {e^{ - 2{\gamma _2}kd}}{r_{21}}{r_{23}};\\
{f^{TM}} = \\
 = 1 - {e^{ - 2{\gamma _2}kd}}\left[ {\frac{{{{\tilde \varepsilon }_1}{\gamma _2} - {{\tilde \varepsilon }_2}{\gamma _1} + 2{\gamma _1}{\gamma _2}\alpha _L^\parallel }}{{{{\tilde \varepsilon }_1}{\gamma _2} + {{\tilde \varepsilon }_2}{\gamma _1} + 2{\gamma _1}{\gamma _2}\alpha _L^\parallel }}} \right]\left[ {\frac{{{{\tilde \varepsilon }_3}{\gamma _2} - {{\tilde \varepsilon }_2}{\gamma _3} + 2{\gamma _2}{\gamma _3}\alpha _R^\parallel }}{{{{\tilde \varepsilon }_3}{\gamma _2} + {{\tilde \varepsilon }_2}{\gamma _3} + 2{\gamma _2}{\gamma _3}\alpha _R^\parallel }}} \right];\\
{f^{TE}} = \\
 = 1 - {e^{ - 2{\gamma _2}kd}}\left[ {\frac{{{\gamma _2} - {\gamma _1} + 2{{\left( {\omega /ck} \right)}^2}\alpha _L^ \bot }}{{{\gamma _2} + {\gamma _1} - 2{{\left( {\omega /ck} \right)}^2}\alpha _L^ \bot }}} \right]\left[ {\frac{{{\gamma _2} - {\gamma _3} + 2{{\left( {\omega /ck} \right)}^2}\alpha _R^ \bot }}{{{\gamma _2} + {\gamma _3} - 2{{\left( {\omega /ck} \right)}^2}\alpha _R^ \bot }}} \right],
\end{array}
\label{equ15}
\end{equation}
where we assumed there are 2D sheets at both interfaces. These sheets may be different so we have put the subscripts $L$ and $R$ on the polarizability of the left and right sheet, respectively.

\section{Dielectric function of graphene\label{dielfunc}}
In terms of the polarizability the dielectric function of a 2D system is given by
$\varepsilon \left( {{\bf{k}},\omega } \right) = 1 + \alpha \left( {{\bf{k}},\omega } \right) = 1 - {v^{2D}}\left( k \right)\chi \left( {{\bf{k}},\omega } \right)$, where ${v^{2D}}\left( k\right) = 2\pi {e^2}/k$ is the 2D fourier transform of the coulomb potential and $\chi \left( {{\bf{k}},\omega } \right)$ the density-density correlation function or polarization bubble. It can be useful to have the dielectric function also expressed in terms of the dynamical conductivity, $\sigma \left( {{\bf{k}},\omega } \right)$. The relation is: $\varepsilon \left( {{\bf{k}},\omega } \right) = 1 + 2\pi i\sigma \left( {{\bf{k}},\omega } \right)k/\omega $. These relations are valid for all 2D systems. 

From now on we limit the treatment to graphene. We discuss the T\,=\,0K functions, only, and argue later that we may use these functions also at 300K. All functions are longitudinal versions; transverse versions are not available in the literature. All functions are derived under the assumption that the band structure for graphene consists of two pair of cones; each pair consists of a conical valence band with the tip in positive energy direction and a conical conduction band with the tip in the negative energy direction; the cones are stacked upon each other so that the two tips coincide; the cones extend from minus to plus infinity. In a real graphene system there are deviations from the conical shapes and there are also other bands contributing. Thus the functions we discuss here are only strictly valid for small frequencies. We estimate the results to be reliable for $\hbar \omega  < 4eV$.

Let us first begin with an undoped graphene sheet. In a general point, $z$, in the complex frequency plane, away from the real axis the density-density correlation function is\cite{Guinea}
\begin{equation}
\chi \left( {{\bf{k}},z} \right) =  - \frac{g}{{16\hbar }}\frac{{k^2 }}{{\sqrt {v^2 k^2  - z^2 } }},
\label{equ16}
\end{equation}
where $v$ is the carrier velocity which is a constant in graphene ($E =  \pm \hbar vk$), and $g$ represents the degeneracy parameter with the value of 4 (a factor of 2 for spin and a factor of 2 for the cone degeneracy.) In the numerical calculations we use the value\,\cite{Wun}  $8.73723 \times {10^5}$ m/s for $v$.

When the graphene sheet is doped the dielectric function becomes much more complicated. However it has been derived by several groups\cite{Ser1,Wun,Sarma}. 
In the two next equations we use dimension-less variables: $x = k/2{k_F}$; $y = \hbar \omega {\rm{ }}/2{E_F}$; $\tilde z = \hbar z/2{E_F}$.  
 The density-density correlation function in a general point in the complex frequency plane, $z$, away from the real axis is\,\cite{Ser1}
\begin{equation}
\begin{array}{*{20}l}
   {\chi \left( {{\bf{k}},z} \right) =  - D_0 \left\{ {1 + \frac{{x^2 }}{{4\sqrt {x^2  - \tilde z^2 } }}\left[ {\pi  - f\left( {x,\tilde z} \right)} \right]} \right\};}  \\
   {f\left( {x,\tilde z} \right) = {\rm{asin}}\left( {\frac{{1 - \tilde z}}{x}} \right) + {\rm{asin}}\left( {\frac{{1 + \tilde z}}{x}} \right)}  \\
   {\qquad \qquad  - \frac{{\tilde z - 1}}{x}\sqrt {1 - \left( {\frac{{\tilde z - 1}}{x}} \right)^2 }  + \frac{{\tilde z + 1}}{x}\sqrt {1 - \left( {\frac{{\tilde z + 1}}{x}} \right)^2 },}  \\
\end{array}
\label{equ17}
\end{equation}
where ${D_0} = \sqrt {gn/\pi {\hbar ^2}{v^2}} $ is the density of states at the fermi level and $n$ is the doping concentration. 
The same result holds for excess of electrons and excess of holes. The expression in Eq.\,(\ref{equ17}) reproduces the results of Refs.\,\onlinecite{Sarma,Wun} when $z$ approaches the real axis from above, apart from an anomaly in the result of Ref.\,\onlinecite{Sarma}. The expression in their Eq.\,(8), which should be purely real has an imaginary part. If this is removed their result agrees with both ours and that in Ref.\,\onlinecite{Wun}. %

In the calculations we need the function at the imaginary frequency axis. We have derived a very useful analytical expression valid along the imaginary axis; an expression in terms of real valued functions of real valued variables:\,\cite{Ser1}
\begin{equation}
\begin{array}{l}
 \chi '\left( {{\bf{k}},\omega } \right) = \chi \left( {{\bf{k}},i\omega } \right) \\ 
 \quad \quad \quad \quad  =  - D_0 \left\{ {1 + \frac{{x^2 }}{{4\sqrt {y^2  + x^2 } }}\left[ {\pi  - g\left( {x,y} \right)} \right]} \right\}; \\ 
 \quad g(x,y) = {\rm{atan}}\left[ {h(x,y)k\left( {x,y} \right)} \right] + l\left( {x,y} \right); \\ 
 \quad h\left( {x,y} \right) = \frac{{2\left\{ {\left[ {x^2 \left( {y^2  - 1} \right) + \left( {y^2  + 1} \right)^2 } \right]^2  + \left( {2yx^2 } \right)^2 } \right\}^{{1 \mathord{\left/
 {\vphantom {1 4}} \right.
 \kern-\nulldelimiterspace} 4}} }}{{\sqrt {\left( {x^2  + y^2  - 1} \right)^2  + \left( {2y} \right)^2 }  - \left( {y^2  + 1} \right)}}, \\ 
 \quad k\left( {x,y} \right) = \sin \left\{ {\frac{1}{2}{\rm{atan}}\left[ {\frac{{2yx^2 }}{{x^2 \left( {y^2  - 1} \right) + \left( {y^2  + 1} \right)^2 }}} \right]} \right\}, \\ 
 \quad \; l\left( {x,y} \right) =  \\ 
 \frac{{\sqrt { - 2x^2 \left( {y^2  - 1} \right) - 2\left( {y^4  - 6y^2  + 1} \right) + 2\left( {y^2  + 1} \right)\sqrt {x^4  + 2x^2 \left( {y^2  - 1} \right) + \left( {y^2  + 1} \right)^2 } } }}{{x^2 }}, \\ 
 \end{array}
 \label{equ18}
\end{equation}
where the arcus tangens function is taken from the branch where $ 0 \le {\rm{atan}} < \pi $. The density-density correlation function on the imaginary frequency axis has been derived before in a compact and inexplicit form (see Ref.\,\onlinecite{Barlas} and references therein.) Here we have chosen to express it in an explicit form in terms of real valued functions of real valued variables.

\section{free-standing graphene sheets\label{twosheets}}
Two free-standing graphene sheets we obtain from our geometry by letting all three media be vacuum and let both sheets have the same polarizability. We further assume that there is negligible difference between transverse and longitudinal screening in graphene. Then the mode condition functions in Eq.\,(\ref{equ15}) reduce into
\begin{equation}
\begin{array}{l}
{f^{TM}} = 1 - {e^{ - 2{\gamma ^{\left( 0 \right)}}kd}}{\left[ {\frac{{{\gamma ^{\left( 0 \right)}}\alpha \left( {k,\omega } \right)}}{{1 + {\gamma ^{\left( 0 \right)}}\alpha \left( {k,\omega } \right)}}} \right]^2};\\
{f^{TE}} = 1 - {e^{ - 2{\gamma ^{\left( 0 \right)}}kd}}{\left[ {\frac{{{{\left( {\omega /ck} \right)}^2}\alpha \left( {k,\omega } \right)}}{{{\gamma ^{\left( 0 \right)}} - {{\left( {\omega /ck} \right)}^2}\alpha \left( {k,\omega } \right)}}} \right]^2},
\end{array}
\label{equ19}
\end{equation}
where
\begin{equation}
{\gamma ^{\left( 0 \right)}}\left( {k,\omega } \right) = \sqrt {1 - {{\left( {\omega /ck} \right)}^2}} .
\label{equ20}
\end{equation}

If retardation effects are neglected there are no TE modes and the mode condition function for TM modes is reduced further into
\begin{equation}
{f^{TM}} = 1 - {e^{ - 2kd}}{\left[ {\frac{{\alpha \left( {k,\omega } \right)}}{{1 + \alpha \left( {k,\omega } \right)}}} \right]^2}.
\label{equ21}
\end{equation}
Now we have all we need for writing down the expression for the interaction energy between two free-standing graphene sheets. It is
%
\begin{figure}
\includegraphics[width=8.0cm]{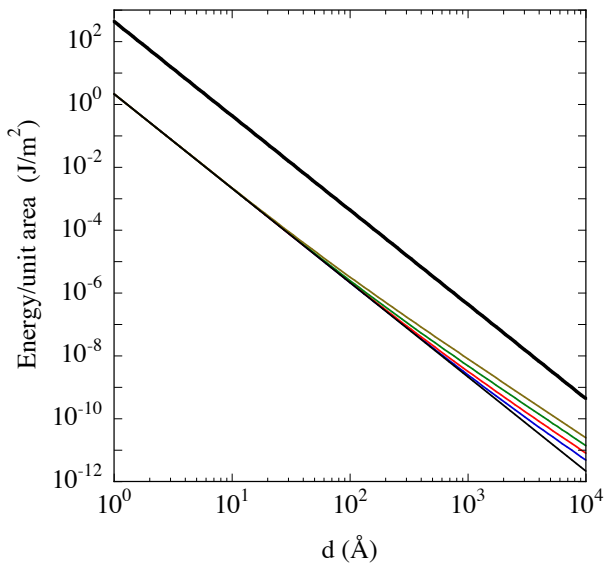}
\caption{(Color online) The attractive non-retarded\cite{Ser1} interaction energy between two graphene sheets. The lower straight line is for undoped sheets, while the bent curves are for doping densities $1 \times 10^{10}, {\rm{ }}1 \times 10^{11}, {\rm{ }}1 \times 10^{12}, {\rm{ and }} {\kern 1pt} 1 \times 10^{13} {\rm{ cm }}^{-2} $, respectively, counted from below. The upper thick straight line is the Casimir classical result for two ideal metal half-spaces.}
\label{figu1}
\end{figure}
\begin{equation}
\begin{array}{l}
E = \hbar \int {\frac{{{d^2}k}}{{{{\left( {2\pi } \right)}^2}}}} \int\limits_0^\infty  {\frac{{d\omega }}{{2\pi }}} \ln \left\{ {1 - {e^{ - 2{\gamma ^{\left( 0 \right)}}kd}}{{\left[ {\frac{{{\gamma ^{\left( 0 \right)}}\alpha \left( {k,i\omega } \right)}}{{1 + {\gamma ^{\left( 0 \right)}}\alpha \left( {k,i\omega } \right)}}} \right]}^2}} \right\}\\
 + \hbar \int {\frac{{{d^2}k}}{{{{\left( {2\pi } \right)}^2}}}} \int\limits_0^\infty  {\frac{{d\omega }}{{2\pi }}} \ln \left\{ {1 - {e^{ - 2{\gamma ^{\left( 0 \right)}}kd}}{{\left[ {\frac{{ - {{\left( {\omega /ck} \right)}^2}\alpha \left( {k,i\omega } \right)}}{{{\gamma ^{\left( 0 \right)}} + {{\left( {\omega /ck} \right)}^2}\alpha \left( {k,i\omega } \right)}}} \right]}^2}} \right\},
\end{array}
\label{equ22}
\end{equation}
in the full retarded treatment. To save space we have suppressed the arguments of the ${\gamma ^{\left( 0 \right)}}\left( {k,i\omega } \right)$ function.
In the non-retarded treatment the energy reduces into
\begin{equation}
E = \hbar \int {\frac{{{d^2}k}}{{{{\left( {2\pi } \right)}^2}}}} \int\limits_0^\infty  {\frac{{d\omega }}{{2\pi }}} \ln \left\{ {1 - {e^{ - 2kd}}{{\left[ {\frac{{\alpha \left( {k,i\omega } \right)}}{{1 + \alpha \left( {k,i\omega } \right)}}} \right]}^2}} \right\}.
\label{equ23}
\end{equation}

At finite temperature the frequency integration in Eqs.(\,\ref{equ22}) and (\ref{equ23}) is replaced by a discrete frequency summation as described in Eq.(\,\ref{equ2}).
\subsection{Undoped graphene sheets\label{twoundopedsheets}}
%
%
\begin{figure}
\includegraphics[width=8.0cm]{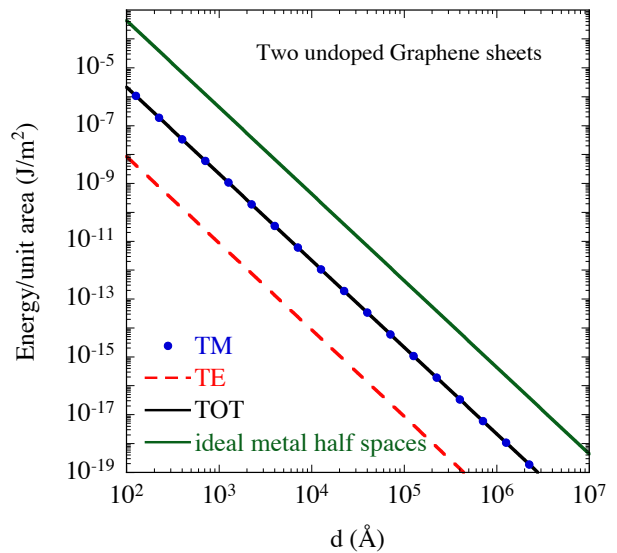}
\caption{(Color online)  The attractive retarded interaction energy between two undoped graphene sheets. The dashed straight line is the TE contribution while the filled circles are the TM contribution. The lower solid straight line is the total contribution, clearly dominated by the TM contribution. The upper thick straight line is the Casimir classical result for two ideal metal half-spaces.}
\label{figu2}
\end{figure}
%
\begin{figure}
\includegraphics[width=8.0cm]{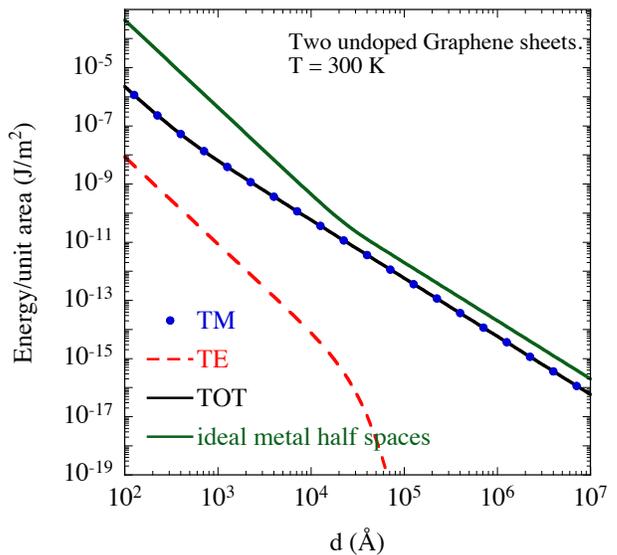}
\caption{(Color online) Same as Fig.\,\ref{figu2} but now at 300 K.}
\label{figu3}
\end{figure}

Let us first discuss two undoped graphene sheets. With the particular screening in graphene it turns out that $\alpha \left( {k/\lambda ,i\omega /\lambda } \right) = \alpha \left( {k,i\omega } \right)$ and the separation dependence of the non-retarded interaction in Eq.\,(\ref{equ23}) becomes very simple. A change in dummy variables removes the only $d$ in the integrand and produces the factor of ${d^{ - 3}}$ in front of the integral. The result is shown as the lower thin straight line in Fig.\,\ref{figu1}. The upper thick straight line is the Casimir classical result for two ideal metal half-spaces. The remaining curves in Fig.\,\ref{figu1} are the results from doped graphene sheets with doping concentrations ${10^{10}}$, ${10^{11}}$, ${10^{12}}$, and ${10^{13}}\,{\rm{c}}{{\rm{m}}^{ - 2}}$, respectively, counting from below. These curves approach asymptotes with $ \sim {d^{ - 5/2}}$ for large separations. These asymptotes will cross the ideal metal half-space result and in the doping case there are bound to be retardation effects. It turns out that also ${\gamma ^{\left( 0 \right)}}\left( {k/\lambda ,i\omega /\lambda } \right) = {\gamma ^{\left( 0 \right)}}\left( {k,i\omega } \right)$ so the separation dependence of the retarded interaction in Eq.\,(\ref{equ22}) becomes equally simple. The same dummy variable change removes the only $d$ in both integrands and produces the factor of ${d^{ - 3}}$ in front of the integrals. In the undoped case the retardation effects are negligible. There is a small reduction in strength of less than one percent.  The retarded results are shown in Fig.\,\ref{figu2}. Note that here the figure begins at 100 \AA. The TE contribution is negligible. If plotted in this figure the non-retarded result would fall within the thickness of the straight line representing the total result. Note that all four curves follow the same power law.

At finite temperature we can no longer use any dummy variable substitution in the frequency variable, so we will not have a simple power law for all distances. Furthermore we need the polarizabilities at finite temperature.  The change in polarizability with temperature is largest for small frequencies. At 300K the static value increases with approximately 10 percent\,\cite{Woods}. The TE amplitude reflection coefficient vanishes at zero frequency and the TM amplitude reflection coefficient increases by approximately one percent due to the finite temperature. The finite temperature correction is dominated by the $n=0$ term in the Matsubara summation that only depends on the static value. Thus we may safely use the zero temperature form of the polarizability. The results are shown in Fig.\,\ref{figu3}. The TE contribution becomes even less important at 300K. The TM contribution approaches a new asymptote $ \sim {d^{ - 2}}$. Also the Casimir classical ideal metal half-space result follows an asymptote with this slope at large enough separations. Note that the asymptote for the total result is somewhat smaller than half the ideal metal asymptote. It would be one half of the ideal metal result if the static polarizability were infinite. Using the approximation of a constant conductivity as in Ref.\,\onlinecite{Woods} would give that result.
\subsection{Doped graphene sheets\label{twodopedsheets}}
\begin{figure}
\includegraphics[width=8.0cm]{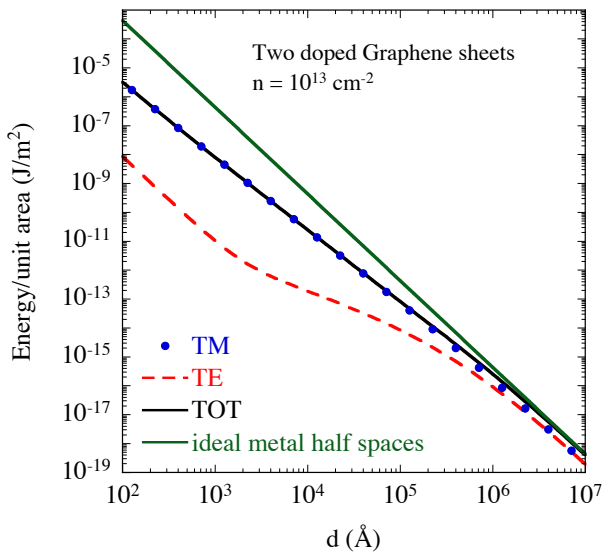}
\caption{(Color online) Same as Fig.\,\ref{figu2} but now for doped sheets.}
\label{figu4}
\end{figure}

When the graphene sheets are doped the dielectric function becomes much more complicated (see Eq.\,(\ref{equ18}).) It is no longer possible to use the simple change of dummy variables that we did before to find a simple power law for the interaction. We limit the presentation to the highest of the four doping concentrations in Fig.\,\ref{figu1}. The results at zero temperature are shown in Fig.\,\ref{figu4}. Here we see that neither the TM nor the TE contribution are straight lines parallel with the ideal metal result. Both increases so that the total result approaches the ideal case. To be noted is that the TE contribution now is as important as the TM contribution. They are responsible for half the result each at large separations. 
%
\begin{figure}
\includegraphics[width=8.0cm]{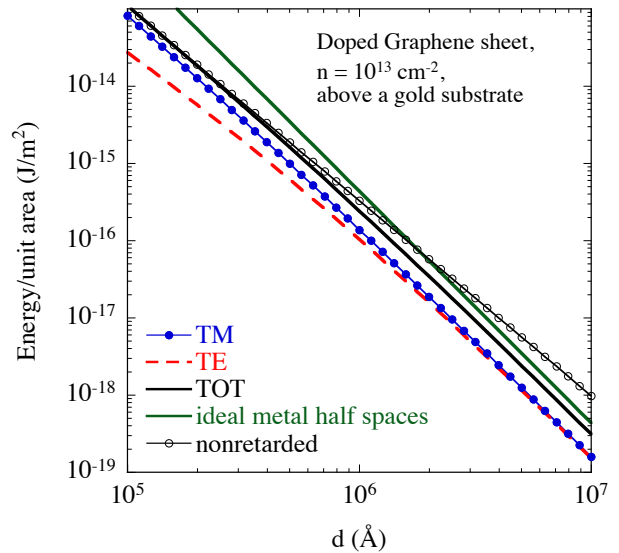}
\caption{(Color online) Expanded view of Fig.\,\ref{figu4} where also the non-retarded result has been added (the thin solid curve with open circles.)}
\label{figu5}
\end{figure}
The total result does not cross the ideal metal half-space result; it stays below but close. This is due to retardation effects. Here it is interesting to compare to the non-retarded results. We show this comparison in an expanded part of the figure as Fig.\,\ref{figu5}. Here we see even clearer that the TE and TM contribute equal amounts. The thin solid curve with open circles is the non-retarded result. It follows the total retarded result at the left end of the figure. Then it detaches itself and crosses the classical Casimir result. In this region the retardation effects are clearly present. 
\begin{figure}
\includegraphics[width=8.0cm]{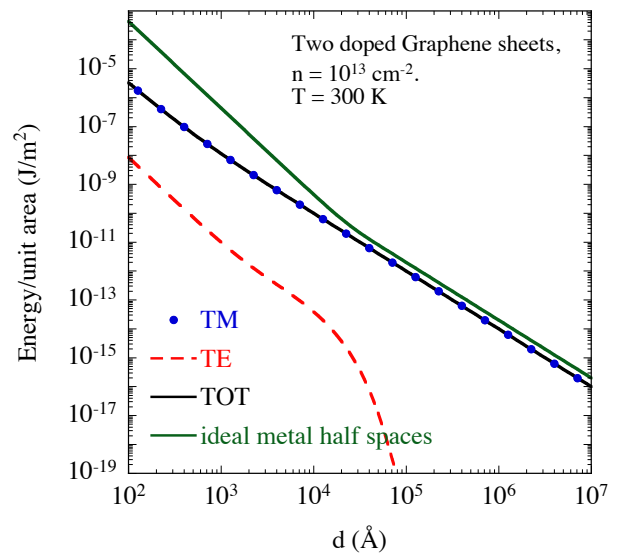}
\caption{(Color online) Same as Fig.\,\ref{figu4} but now at T = 300\,K.}
\label{figu6}
\end{figure}

 At T = 300\,K the TE result decreases in strength and vanishes for large separations. The results are shown in Fig.\,\ref{figu6}. The TM contribution then dominates completely for large separations. The TM waves result in a contribution half the strength of the total contribution in the ideal Casimir geometry. This is because the static polarizability now diverges resulting in a TM reflection coefficient equal to unity. This behavior is completely analogous to the result in the real metal half-spaces geometry when the experimentally obtained dielectric functions are used or the Drude model. Then the absence of TE contribution is due to dissipation. Here it is due to spatial dispersion. One of the benefits of geometries consisting of 2D sheets from the fundamental physics point of view is that spatial dispersion (momentum dependence of the dielectric function) is automatically fully included.  To include spatial dispersion in a system of thick layers is very complicated\,\cite{Ser3}. In the large separation limit only the classical ($n = 0$) result survives. According to the Bohr -- van Leeuwen theorem\,\cite{Bim} there should be no classical TE contribution. Thus the theorem is fulfilled also in the present system.
\section{Graphene sheet above a gold substrate\label{gold}}
A free-standing graphene sheet above a substrate we obtain from our geometry by letting media $1$ and $2$ be vacuum and medium $3$ be the substrate. We further assume that there is negligible difference between transverse and longitudinal screening in graphene. Then the mode condition functions in Eq.\,(\ref{equ15}) reduce into
\begin{equation}
\begin{array}{*{20}{l}}
{{f^{TM}} = 1 - {e^{ - 2{\gamma ^{\left( 0 \right)}}kd}}\left[ {\frac{{{\gamma ^{\left( 0 \right)}}\alpha \left( {k,\omega } \right)}}{{1 + {\gamma ^{\left( 0 \right)}}\alpha \left( {k,\omega } \right)}}} \right]\left[ {\frac{{{{\tilde \varepsilon }_s}\left( \omega  \right){\gamma ^{\left( 0 \right)}} - {\gamma _s}}}{{{{\tilde \varepsilon }_s}\left( \omega  \right){\gamma ^{\left( 0 \right)}} + {\gamma _s}}}} \right];}\\
{{f^{TE}} = 1 - {e^{ - 2{\gamma ^{\left( 0 \right)}}kd}}\left[ {\frac{{{{\left( {\omega /ck} \right)}^2}\alpha \left( {k,\omega } \right)}}{{{\gamma ^{\left( 0 \right)}} - {{\left( {\omega /ck} \right)}^2}\alpha \left( {k,\omega } \right)}}} \right]\left[ {\frac{{{\gamma ^{\left( 0 \right)}} - {\gamma _s}}}{{{\gamma ^{\left( 0 \right)}} + {\gamma _s}}}} \right].}
\end{array}
\label{equ24}
\end{equation}
If retardation effects are neglected there are no TE modes and the mode condition function for TM modes is reduced further into
\begin{equation}
{f^{TM}} = 1 - {e^{ - 2kd}}\left[ {\frac{{\alpha \left( {k,\omega } \right)}}{{1 + \alpha \left( {k,\omega } \right)}}} \right]\left[ {\frac{{{{\tilde \varepsilon }_s}\left( \omega  \right) - 1}}{{{{\tilde \varepsilon }_s}\left( \omega  \right) + 1}}} \right].
\label{equ25}
\end{equation}
\begin{figure}
\includegraphics[width=8.0cm]{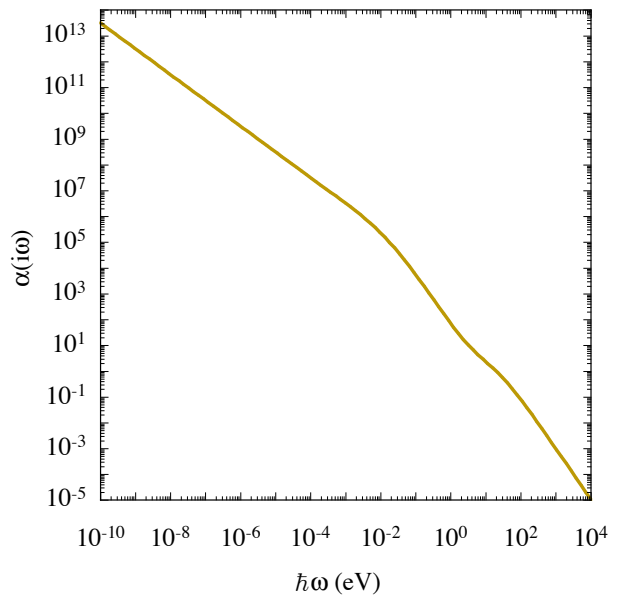}
\caption{(Color online) The polarizability for gold at the imaginary axis used in the numerical calculations.}
\label{Figu7}
\end{figure}
%
\begin{figure}
\includegraphics[width=8.0cm]{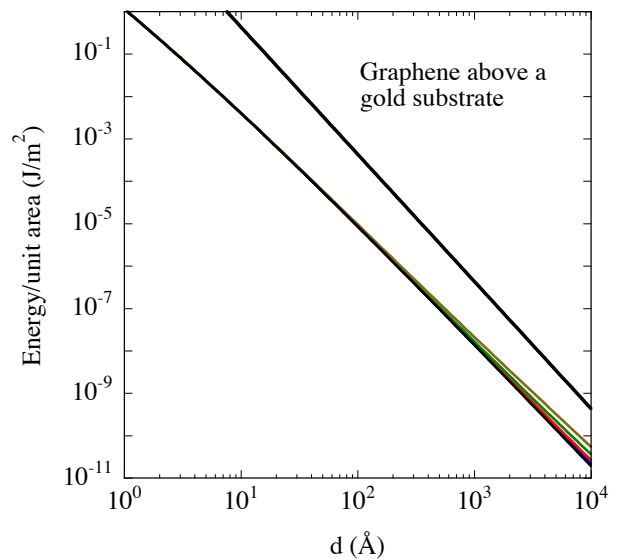}
\caption{(Color online) The attractive non-retarded\cite{Ser1} interaction energy between a graphene sheet and a gold substrate. The lowest curve is for an undoped graphene sheet  while the other curves are for doping densities $1 \times 10^{10}, {\rm{ }}1 \times 10^{11}, {\rm{ }}1 \times 10^{12}, {\rm{ and }} {\kern 1pt} 1 \times 10^{13} {\rm{ cm }}^{-2} $, respectively, counted from below. The upper thick straight line is the Casimir classical result for two ideal metal half-spaces.}
\label{Figu8}
\end{figure}
Now we have all we need for writing down the expression for the interaction energy for a free-standing graphene sheet above a substrate. It is
\begin{equation}
\begin{array}{*{20}{l}}
{\begin{array}{*{20}{l}}
{E = \hbar \int {\frac{{{d^2}k}}{{{{\left( {2\pi } \right)}^2}}}} \int\limits_0^\infty  {\frac{{d\omega }}{{2\pi }}} \ln \left\{ {1 - {e^{ - 2{\gamma ^{\left( 0 \right)}}kd}}\left[ {\frac{{{\gamma ^{\left( 0 \right)}}\alpha \left( {k,i\omega } \right)}}{{1 + {\gamma ^{\left( 0 \right)}}\alpha \left( {k,i\omega } \right)}}} \right]} \right.}\\
{\quad \quad \quad \quad \quad \quad \quad \quad \quad \quad \quad \quad \quad \quad  \times \left. {\left[ {\frac{{{{\tilde \varepsilon }_s}\left( {i\omega } \right){\gamma ^{\left( 0 \right)}} - {\gamma _s}}}{{{{\tilde \varepsilon }_s}\left( {i\omega } \right){\gamma ^{\left( 0 \right)}} + {\gamma _s}}}} \right]} \right\}}
\end{array}}\\
\begin{array}{*{20}{l}}
{ + \hbar \int {\frac{{{d^2}k}}{{{{\left( {2\pi } \right)}^2}}}} \int\limits_0^\infty  {\frac{{d\omega }}{{2\pi }}} \ln \left\{ {1 - {e^{ - 2{\gamma ^{\left( 0 \right)}}kd}}\left[ {\frac{{ - {{\left( {\omega /ck} \right)}^2}\alpha \left( {k,i\omega } \right)}}{{{\gamma ^{\left( 0 \right)}} + {{\left( {\omega /ck} \right)}^2}\alpha \left( {k,i\omega } \right)}}} \right]} \right.}\\
{\left. {\quad \quad \quad \quad \quad \quad \quad \quad \quad \quad \quad \quad \quad \quad \quad \quad \quad  \times \left[ {\frac{{{\gamma ^{\left( 0 \right)}} - {\gamma _s}}}{{{\gamma ^{\left( 0 \right)}} + {\gamma _s}}}} \right]} \right\},}
\end{array}
\end{array}
\label{equ26}
\end{equation}
in the full retarded treatment. To save space we have suppressed the arguments of the ${\gamma ^{\left( 0 \right)}}\left( {k,i\omega } \right)$ and ${\gamma _s}\left( {k,i\omega } \right)$ functions.
In the non-retarded treatment the energy reduces into
\begin{equation}
\begin{array}{*{20}{l}}
{E = \hbar \int {\frac{{{d^2}k}}{{{{\left( {2\pi } \right)}^2}}}} \int\limits_0^\infty  {\frac{{d\omega }}{{2\pi }}} \ln \left\{ {1 - {e^{ - 2kd}}\left[ {\frac{{\alpha \left( {k,i\omega } \right)}}{{1 + \alpha \left( {k,i\omega } \right)}}} \right]} \right.}\\
{\quad \quad \quad \quad \quad \quad \quad \quad \quad \quad \quad  \times \left. {\left[ {\frac{{{{\tilde \varepsilon }_s}\left( {i\omega } \right) - 1}}{{{{\tilde \varepsilon }_s}\left( {i\omega } \right) + 1}}} \right]} \right\}.}
\end{array}
\label{equ27}
\end{equation}

At finite temperature the frequency integration in Eqs.(\,\ref{equ26}) and (\ref{equ27}) is replaced by a discrete frequency summation as described in Eq.(\,\ref{equ2}).

Before we continue we need to discuss the gold dielectric function, ${{{\tilde \varepsilon }_s}\left( {i\omega } \right)}$. We have used experimental optical data\,\cite{Palik} for the imaginary part of the dielectric function. We have extrapolated the data both for large and small frequencies and used a version\,\cite{Ser0} of the Kramers Kronig dispersion relations to find the result at the imaginary axis. The resulting data are shown in Fig.\,\ref{Figu7}.

\subsection{Undoped graphene sheet\label{substrateundoped}}
%
\begin{figure}
\includegraphics[width=8.0cm]{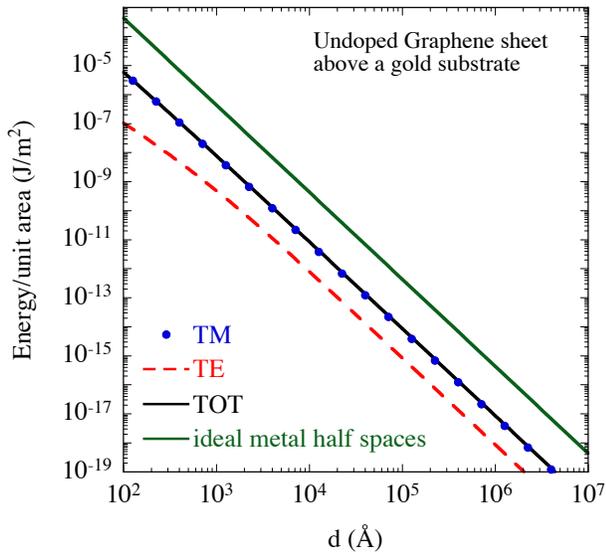}
\caption{(Color online) The attractive retarded interaction energy between a graphene sheet and a gold substrate. The dashed curve is the TE contribution and the filled circles are the TM contribution. The lower solid curve is the total contribution. The upper thick straight line is the Casimir classical result for two ideal metal half-spaces.}
\label{Figu9}
\end{figure}
The non-retarded results are shown in Fig.\,\ref{Figu8}. The lowest thin curve is the result for an undoped graphene sheet above a gold substrate. We note that the result no longer follows the simple power law that we found for two free-standing graphene sheets. However, for large separations the curve approaches an asymptote $ \sim {d^{ - 3}}$ and is parallel to the classical Casimir result (upper thick straight line). Included in the figure are also the results for doped graphene sheets. These curves are for doping densities $1 \times 10^{10}, {\rm{ }}1 \times 10^{11}, {\rm{ }}1 \times 10^{12}, {\rm{ and }} {\kern 1pt} 1 \times 10^{13} {\rm{ cm }}^{-2} $, respectively, counted from below. The curve for the undoped graphene sheet will never come close to the classical Casimir result and we do not expect to find any important retardation effects. However, the results for the doped sheets will sooner or later cross the classical result and there should be important retardation effects. 

In the retarded treatment we see from Eq.\,(\ref{equ26}) that in general it is not possible to make the dummy variable change that removes the only $d$ in the integrand and produces the factor of ${d^{ - 3}}$ in front of the integral. However for very large $d$ values the exponential factor in the logarithm forces the frequency (and also $k$) to be small in order for the integrand to be non-negligible. This means that for large $d$ values we may use the static value of the substrate dielectric function. When we do that we may use the dummy variable change and both the TM and TE contributions follow the power law ${d^{ - 3}}$. Thus for general separations the result will not follow a simple power law, but for large separations the result approaches an asymptote with the power law ${d^{ - 3}}$. This is what we find in Fig.\,\ref{Figu9} where
we show the retarded results for an undoped graphene sheet above a gold substrate. Just as in the case of two free-standing undoped graphene sheets the TE contribution is much weaker than the TM contribution. On this plot the non-retarded results would fall on the curve for the total contribution.

\begin{figure}
\includegraphics[width=8.0cm]{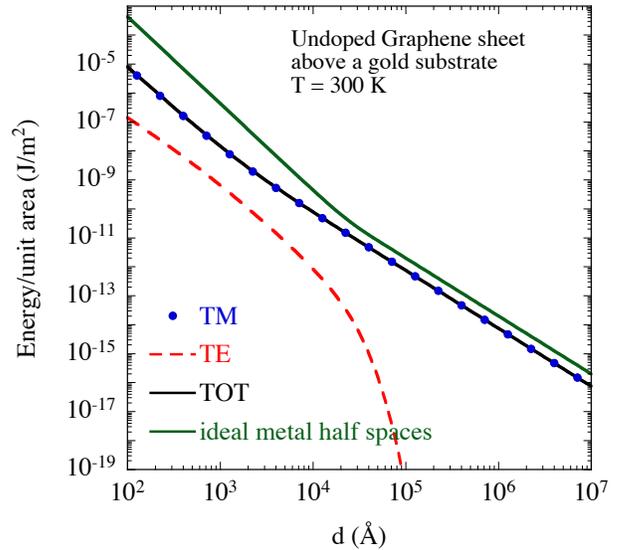}
\caption{(Color online) Same as Fig.\,\ref{Figu9} but now for 300\,K}
\label{Figu10}
\end{figure}

At 300 K (See Fig.\,\ref{Figu10}) the TE contribution becomes even less important. The TM contribution approaches a new asymptote $ \sim {d^{ - 2}}$. Also the Casimir classical ideal metal half-space result follows an asymptote with this slope at large enough separations.
%
\begin{figure}
\includegraphics[width=8.0cm]{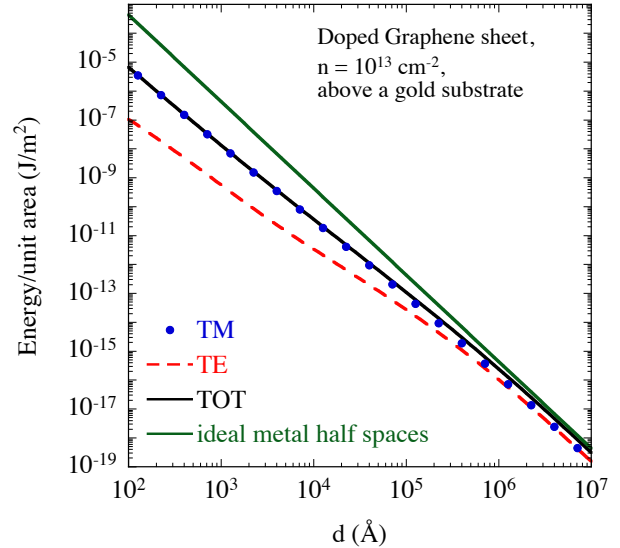}
\caption{(Color online) Same as Fig.\,\ref{Figu9} but now for a doped sheet}
\label{Figu11}
\end{figure}
\subsection{Doped graphene sheet\label{substratedoped}}

\begin{figure}
\includegraphics[width=8.0cm]{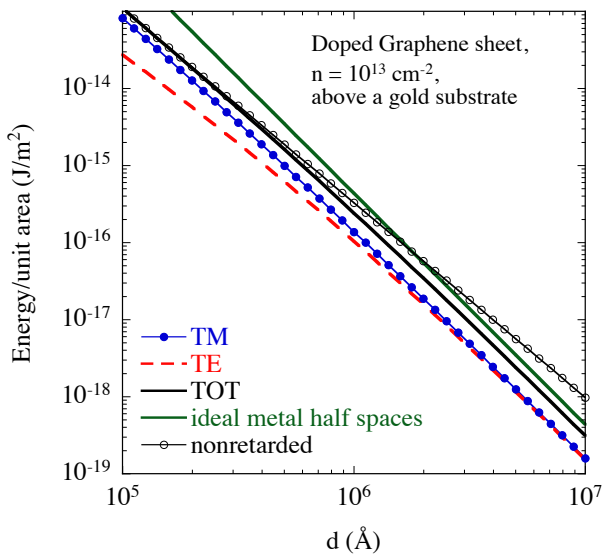}
\caption{(Color online) Expanded view of Fig.\,\ref{Figu11} where also the non-retarded result has been added (the thin solid curve with open circles.)}
\label{Figu12}
\end{figure}

Just as in the case of two free-standing doped graphene sheets the TE contribution becomes important and produces half of the result for large separations. The results are shown in Fig.\,\ref{Figu11}. The total result does not cross the ideal metal half-space result; it stays below but close. This is due to retardation effects. Here it is interesting to compare to the non-retarded results. We show this comparison in an expanded part of the figure in Fig.\,\ref{Figu12}. Here we see even clearer that the TE and TM modes contribute equal amounts. The thin solid curve with open circles is the non-retarded result. It follows the total retarded result at the left end of the figure. Then it detaches itself and crosses the classical Casimir result. In this region the retardation effects are clearly present. 

\begin{figure}
\includegraphics[width=8.0cm]{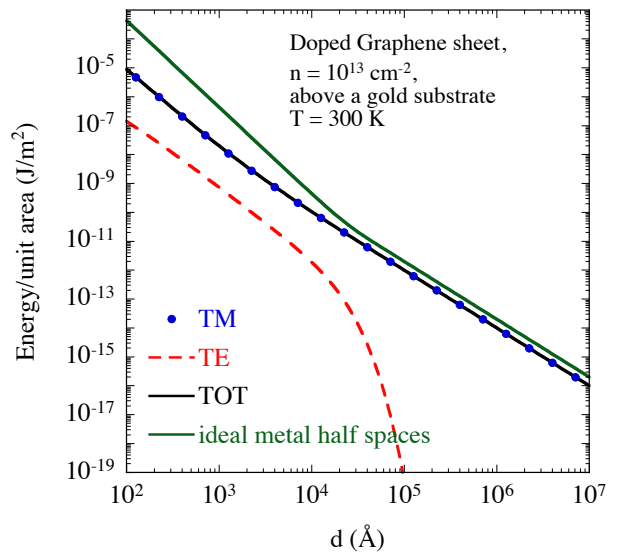}
\caption{(Color online) Same as Fig.\,\ref{Figu11} but now for T = 300\,K }
\label{Figu13}
\end{figure}

At T = 300\,K the TE result decreases in strength and vanishes for large separations. The results are shown in Fig.\,\ref{Figu13}. The TM contribution then dominates completely for large separations. The TM waves result in a contribution with half the strength of the total contribution in the ideal Casimir geometry. The reason is that the static values of the graphene polarizability and the gold dielectric function diverges; the TM reflection coefficients for a vacuum-graphene interface and a vacuum-gold interface are both equal to unity; both TE reflection coefficients are zero because ${\omega ^2}\alpha \left( {k,i\omega } \right)$ and ${\omega ^2}{{\tilde \varepsilon }_s}\left( {i\omega } \right)$ go to zero when the frequency goes towards zero.

This is completely analogous both to the result for two doped free-standing graphene sheets and to the result in the real metal half-spaces geometry when the experimentally obtained dielectric functions are used or the Drude model. Here it is due both to spatial dispersion and to dissipation. Also in the present geometry the Bohr -- van Leeuwen theorem is obviously fulfilled.
\section{Graphene sheet above an ideal metal substrate\label{ideal}}
An ideal metal substrate can be used as a reference system. It gives an upper limit for the interaction with real substrates. It is much easier to use than a real substrate since one avoids the complicated derivation of the dielectric function of the substrate on the imaginary frequency axis using Kramers Kronig dispersion relations. We include an ideal metal substrate for these reasons and for an additional reason.  We found\cite{Ser1}, v.i.z., that the non-retarded result at least for an undoped graphene sheet above an ideal metal substrate does not exist. The integrals do no not converge. We here show that there are no convergence problems in the retarded treatment.

The mode condition function for a free-standing graphene sheet above an ideal metal substrate we obtain from the results for a real substrate in Eq.\,(\ref{equ24}) by letting the dielectric function of the substrate go to infinity. Then the TM and TE amplitude reflections coefficients for the vacuum-substrate interface become unity and minus unity, respectively. We find
\begin{equation}
\begin{array}{*{20}{l}}
{{f^{TM}} = 1 - {e^{ - 2{\gamma ^{\left( 0 \right)}}kd}}\left[ {\frac{{{\gamma ^{\left( 0 \right)}}\alpha \left( {k,\omega } \right)}}{{1 + {\gamma ^{\left( 0 \right)}}\alpha \left( {k,\omega } \right)}}} \right];}\\
{{f^{TE}} = 1 - {e^{ - 2{\gamma ^{\left( 0 \right)}}kd}}\left[ {\frac{{ - {{\left( {\omega /ck} \right)}^2}\alpha \left( {k,\omega } \right)}}{{{\gamma ^{\left( 0 \right)}} - {{\left( {\omega /ck} \right)}^2}\alpha \left( {k,\omega } \right)}}} \right].}
\end{array}
\label{equ28}
\end{equation}
If retardation effects are neglected there are no TE modes and the mode condition function for TM modes is reduced further into
\begin{equation}
{f^{TM}} = 1 - {e^{ - 2kd}}\left[ {\frac{{\alpha \left( {k,\omega } \right)}}{{1 + \alpha \left( {k,\omega } \right)}}} \right].
\label{equ29}
\end{equation}

Now we have all we need for writing down the expression for the interaction energy for a free-standing graphene sheet above an ideal metal substrate. It is
\begin{equation}
\begin{array}{*{20}{l}}
{E = \hbar \int {\frac{{{d^2}k}}{{{{\left( {2\pi } \right)}^2}}}} \int\limits_0^\infty  {\frac{{d\omega }}{{2\pi }}} \ln \left\{ {1 - {e^{ - 2{\gamma ^{\left( 0 \right)}}kd}}\left[ {\frac{{{\gamma ^{\left( 0 \right)}}\alpha \left( {k,i\omega } \right)}}{{1 + {\gamma ^{\left( 0 \right)}}\alpha \left( {k,i\omega } \right)}}} \right]} \right\}}\\
{ + \hbar \int {\frac{{{d^2}k}}{{{{\left( {2\pi } \right)}^2}}}} \int\limits_0^\infty  {\frac{{d\omega }}{{2\pi }}} \ln \left\{ {1 - {e^{ - 2{\gamma ^{\left( 0 \right)}}kd}}\left[ {\frac{{{{\left( {\omega /ck} \right)}^2}\alpha \left( {k,i\omega } \right)}}{{{\gamma ^{\left( 0 \right)}} + {{\left( {\omega /ck} \right)}^2}\alpha \left( {k,i\omega } \right)}}} \right]} \right\}},
\end{array}
\label{equ30}
\end{equation}
\begin{figure}
\includegraphics[width=8.0cm]{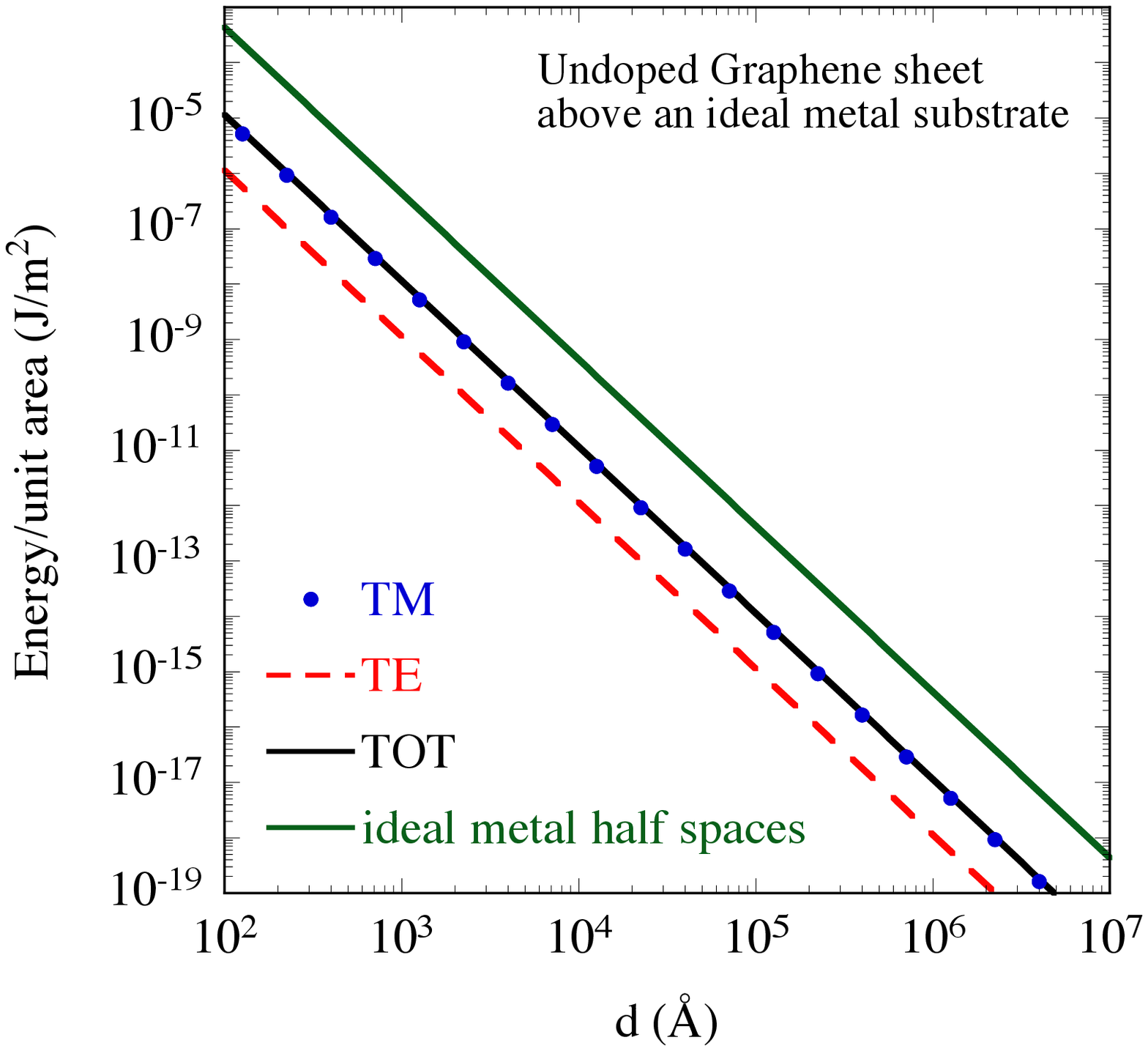}
\caption{(Color online) Same as Fig.\,\ref{Figu9} but now for an ideal metal substrate}
\label{Figu14}
\end{figure}
in the full retarded treatment. To save space we have suppressed the arguments of the ${\gamma ^{\left( 0 \right)}}\left( {k,i\omega } \right)$ function. In the non-retarded treatment the energy reduces into
\begin{equation}
E = \hbar \int {\frac{{{d^2}k}}{{{{\left( {2\pi } \right)}^2}}}} \int\limits_0^\infty  {\frac{{d\omega }}{{2\pi }}} \ln \left\{ {1 - {e^{ - 2kd}}\left[ {\frac{{\alpha \left( {k,i\omega } \right)}}{{1 + \alpha \left( {k,i\omega } \right)}}} \right]} \right\}.
\label{equ31}
\end{equation}
This integral does not converge, at least not in the undoped case. The reason is that the polarizabilty decays too slowly with frequency.

At finite temperature the frequency integration in Eqs.(\,\ref{equ30}) and (\ref{equ31}) is replaced by a discrete frequency summation as described in Eq.(\,\ref{equ2}).
\subsection{Undoped graphene sheet}
In the case of an undoped graphene sheet above an ideal metal substrate we see in Eq.\,(\ref{equ30}) that we may use the dummy variable change in both integrals and both the TM and TE contributions follow the power law ${d^{ - 3}}$. This is verified in Fig.\,\ref{Figu14} where the result for an undoped graphene sheet above an ideal metal substrate is shown. As can be seen the results are very similar to those for an undoped graphene sheet above a gold substrate, Fig.\,\ref{Figu9}. Here we can not compare to the non-retarded result since it does not exist. 

\begin{figure}
\includegraphics[width=8.0cm]{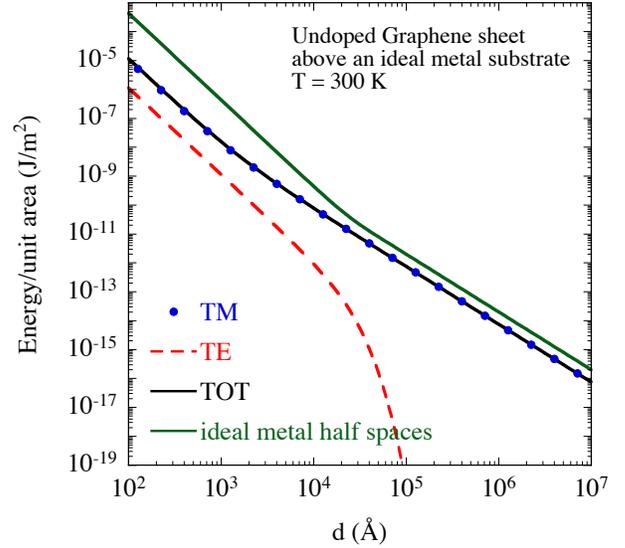}
\caption{(Color online) Same as Fig.\,\ref{Figu10} but now for an ideal metal substrate}
\label{Figu15}
\end{figure}

The results for 300\,K are shown in Fig.\,\ref{Figu15}. Again the results are very close to those for a graphene sheet above a gold substrate, Fig.\,\ref{Figu10}. However, for smaller separations outside the figures the total result for the gold substrate has a smaller slope.
%
\begin{figure}
\includegraphics[width=8.0cm]{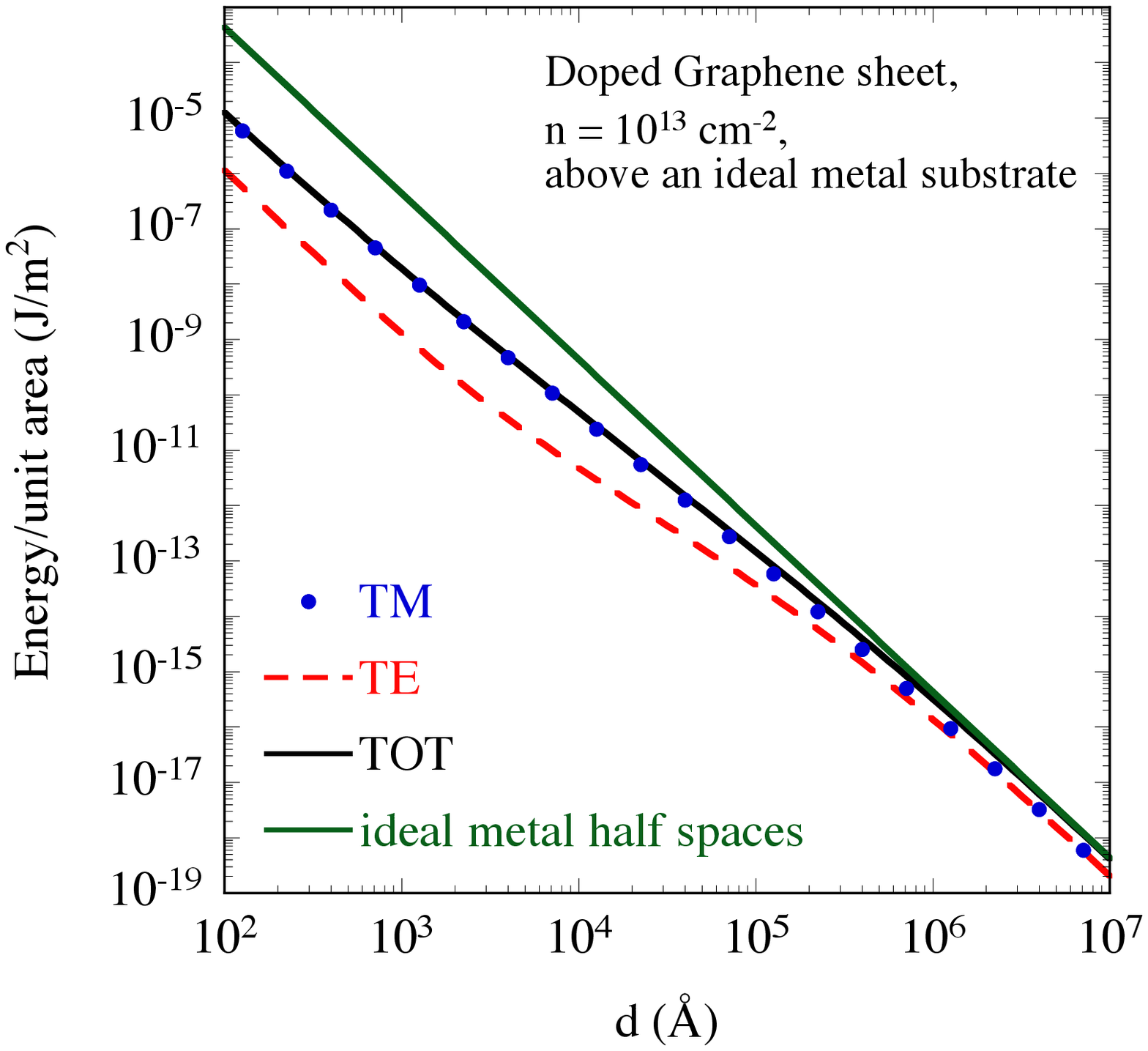}
\caption{(Color online) Same as Fg.\,\ref{Figu11} but now for an ideal metal substrate}
\label{Figu16}
\end{figure}
%
\begin{figure}
\includegraphics[width=8.0cm]{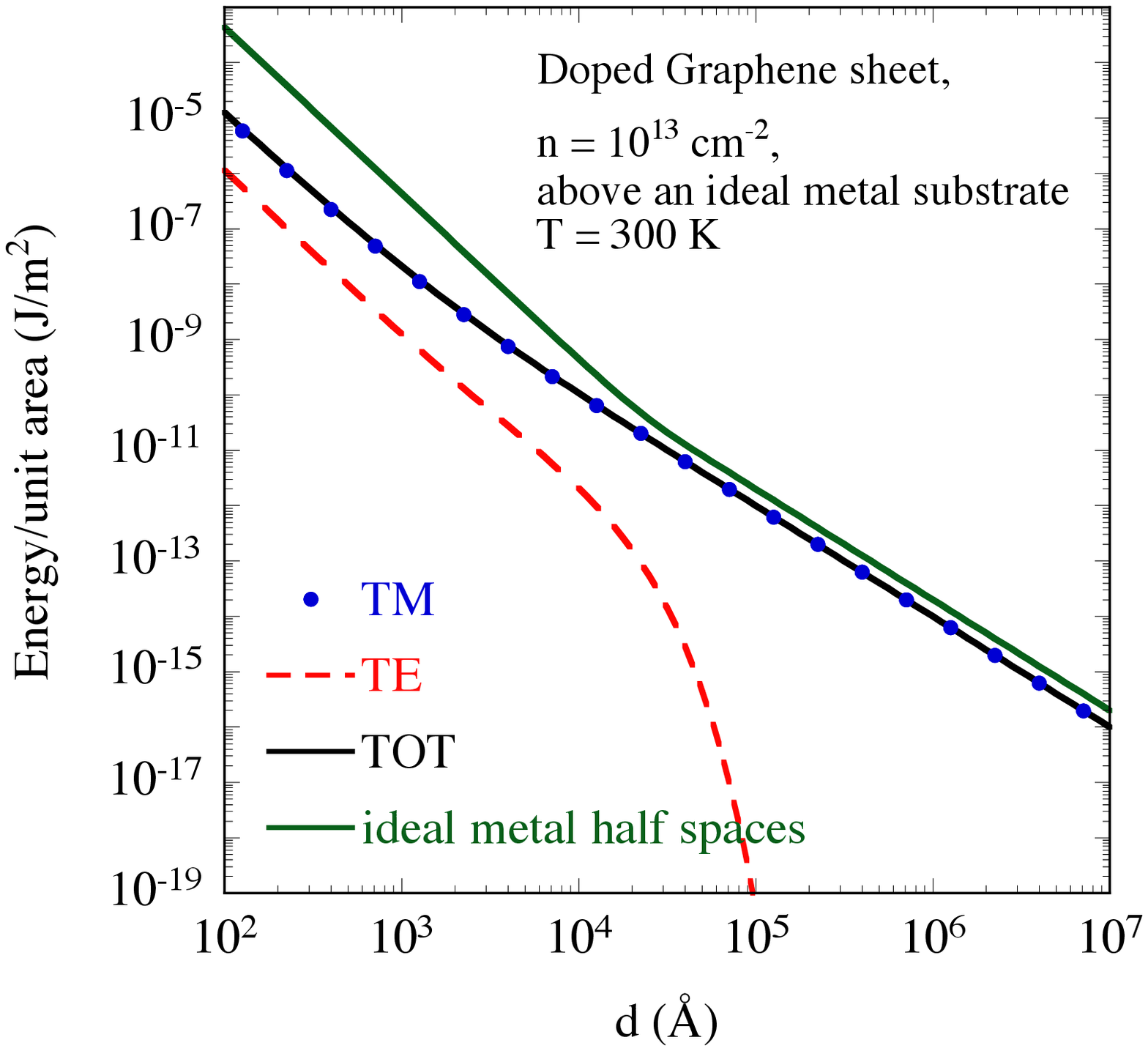}
\caption{(Color online) Same as Fg.\,\ref{Figu13} but now for an ideal metal substrate}
\label{Figu17}
\end{figure}

\subsection{Doped graphene sheet}

The result for a doped graphene sheet above an ideal metal substrate is shown in Fig.\,\ref{Figu16}. As can be noted the results are very close to those for a doped graphene sheet above a gold substrate, Fig.\,\ref{Figu11}. 

The room temperature results, T = 300\,K, are shown in Fig.\,\ref{Figu17}. Again the results are very similar to the ones for a doped graphene sheet above a gold substrate, Fig.\,\ref{Figu13}. However, for smaller separations outside the figures also here the total result for the gold substrate has a smaller slope.


\section{Summary and conclusions\label{summary}}

We have here demonstrated how to incorporate 2D sheets in the formalism for planar structures. We applied the results to graphene systems.  We found that for two free-standing undoped graphene sheets the retardation effects are negligible.  The same holds for an undoped graphene sheet above a gold substrate. However, for doped graphene sheets, two free-standing or one free-standing above a gold substrate, the retardation effects are bound to be important at large enough separations. However, at T = 300\,K the retardation effects again go away. The retarded results for a graphene sheet above an ideal metal substrate are very similar to those for a graphene sheet above a gold substrate. 
On the other hand, the non-retarded results do not exist, so there we have a drastic difference. We should furthermore note that the "saturation effect" at small separations found in the interaction between two 2D metallic sheets\cite{ SerBjo} is absent in geometries involving undoped graphene sheets. This spatial dispersion effect, present also in the case of doped graphene sheets in the region where the doping carriers contribute, is due to the weakening of the screening in the sheet for large momenta. This happens for a momentum where the dispersion curve of the collective mode enters the single particle continuum. In the calculations for T = 300\,K we used the zero temperature polarizabilities for graphene. We estimate this to underestimate the large $d$ asymptote with less than one percent. Furthermore, we used the longitudinal instead of transverse expressions for the graphene polarizabilities in the TE contributions; the reason for this is that nobody has derived the transverse versions, as far as we know; we plan to do this in the future. However, the TE contributions were found to be negligible in most situations.

Retarded results for graphene systems have been obtained by several groups\cite{Bordag2,Woods,Fial, Sara, Svet} before and Santos\cite{Santos} has discussed the unimportance of retardation in graphene systems. We will here go through these results one by one and compare them to ours. 

The interaction between an undoped graphene sheet and an ideal metal substrate was studied in Ref.\,\onlinecite{Bordag2}. In their model they let graphene have a small but finite band gap and the carriers have non-zero mass. Thus, the comparison with our results is meaningless. 

In Ref.\,\onlinecite{Woods} one studied the force between undoped graphene sheets at zero temperature. They used a model where the conductivity is a constant, $\sigma \left( {{\bf{k}},\omega } \right) = {\sigma _0} = {e^2}/4\hbar $. This is the limiting form for $vk <  < \omega$,
\begin{equation}
\begin{array}{l}
\sigma \left( {{\bf{k}},i\omega } \right) = \omega \alpha \left( {{\bf{k}},i\omega } \right)/2\pi k =  - \omega {e^2}\chi \left( {{\bf{k}},i\omega } \right)/{k^2}\\
 = \frac{{\omega {e^2}}}{{{k^2}}}\frac{g}{{16\hbar }}\frac{{{k^2}}}{{\sqrt {{v^2}{k^2} + {\omega ^2}} }} \approx {e^2}/4\hbar ,\,\,\,vk <  < ck < \omega. 
\end{array}
\label{equ32}
\end{equation}
One has to be careful with this approximation, since it is not valid in the finite temperature $n=0$ term, which dominates at large separations.
They gave in their Eq.\,(36), for the first term of a perturbation expansion whereupon ${\sigma _0}$ is small, the expression $E =  - {e^2}/\left( {32\pi {d^3}} \right)$ which is very close to our result. Note that it does not contain any material parameters.  Furthermore they showed in their Fig. 6 how their conductivity varied with temperature; based on this result we concluded that we, to a good approximation, could use the zero temperature expression for the graphene polarizability in our room temperature calculations.

In Ref.\,\onlinecite{Fial} one studied the interaction between an undoped graphene sheet and a substrate (gold and ideal metal) at finite temperature. They used a different approach to the problem starting from a quantum-field-theory model at finite temperature and used different expansions. Our results show an overall agreement. However, in the high temperature limit they put  $v = 0$, which means that they use in that limit the same constant conductivity approximation used in Ref.\,\onlinecite{Woods}. This means that the graphene polarizability diverges at zero frequency and the TM reflection coefficient becomes unity. The result from this is that the interaction in this limit is half the corresponding value between two ideal metal half spaces, just as in the case of a doped graphene sheet above a gold or ideal metal substrate. We find a somewhat smaller value (see Figs.\,\ref{Figu10} and \ref{Figu15}.)

In Ref.\,\onlinecite{Sara} one studied the interaction between undoped and doped graphene sheets at zero temperature and at 300 K; also the interaction between a graphene sheet and a silica substrate was determined. Our results disagree with theirs. For two undoped graphene sheets they find many distance regions, each with a different power law. In the doped case the effect of doping sets in at very small separations with a fractional power law while in our results it sets in at much larger separations; their high temperature asymptote is much weaker than the ideal metal asymptote. We are convinced that their results are wrong. From the first appearance it looks as if they use exactly the same dielectric functions that we do for the undoped and doped graphene. After reading more carefully we believe that they have treated the strictly 2D sheets as layers of finite thickness, using the 2D dielectric functions in place of the bulk dielectric functions of the materials making up the layers. Then they have let the layers have a small but finite thickness. This procedure is not allowed. 

Finally, in Ref.\,\onlinecite{Svet} the focus was on the change in force between a silica membrane and a metal plate when the membrane was covered by a graphene sheet. This work is just vaguely related to ours. We did not find anything that contradicts the findings in our present work.

\begin{acknowledgments}
We are grateful for financial support from the Swedish Research Council, VR-contract No:70529001 
\end{acknowledgments}


\end{document}